# On the recent seismic activity at eastern Aegean Sea: Analysis of fracture-induced electromagnetic emissions in terms of critical fluctuations.

Y. Contoyiannis[1], S. M. Potirakis[1], J. Kopanas[2], G. Antonopoulos[2],
G. Koulouras[3], K. Eftaxias[2], C. Nomicos[3].

1. Department of Electronics Engineering, Piraeus University of Applied Sciences (TEI of Piraeus), 250 Thivon and P. Ralli, Aigaleo, Athens GR-12244, Greece, (Y.C: yiaconto@puas.gr; S.M.P.: spoti@puas.gr)
2. Department of Physics, Section of Solid State Physics, University of Athens, Panepistimiopolis, GR-15784, Zografos, Athens, Greece, (J.K.: jkopan@otenet.gr; G.A.: sv8rx@teiion.gr; K.E: ceftax@phys.uoa.gr)
3. Department of Electronics Engineering, Technological Educational Institute (TEI) of Athens, Ag. Spyridonos, GR-12210, Aigaleo, Athens, Greece, (G.K.: gregkoul@teiath.gr; C.N.: cnomicos@teiath.gr).

**Abstract**

In this letter we show, in terms of fracture-induced electromagnetic emissions (EME) that the Earth system around the focal areas came to critical condition a few days before the occurrence of recent significant (M>5) earthquakes (EQs) which happened in the region of the eastern Aegean Sea, between the Greek Islands of Lesvos (6-7, 12 February 2017 and 12 June 2017) and Kos (20 July 2017) and the Turkish Asia Minor coastline. Moreover, departure from the critical state in terms of a tricritical crossover was identified in the EME recorded prior to the 12 June main event as well as prior to the 20 July main event. The analysis was performed by means of the method of critical fluctuations (MCF).

## 1    Introduction

The possible connection of the electromagnetic (EM) activity that is observed prior to significant earthquakes (EQs) with the corresponding EQ preparation processes, often referred to as seismo-electromagnetics, has been intensively investigated during the last years. Several possible EQ precursors have been suggested in the literature [*Uyeda et al.*, 2009, 2013; *Cicerone et al.*, 2009; *Hayakawa*, 2013a, 2013b, 2015]. The possible relation of the field observed fracture-induced electromagnetic emissions (EME) in the frequency bands of MHz and kHz has been examined in a series of publications [e.g., *Eftaxias et al.*, 2001, 2004, 2008, 2013; *Eftaxias and Potirakis*, 2013; *Contoyiannis et al.,* 2004a, 2005, 2010, 2013, 2015; *Contoyiannis and Eftaxias, 2008*; *Kapiris et al.*, 2004; *Karamanos et al.,* 2006; *Papadimitriou et al.*, 2008; *Potirakis et al.*, 2011, 2012a, 2012b, 2012c, 2013, 2015, 2016; *Minadakis et al.*, 2012a, 2012b; *Donner et al.*, 2015; *Kalimeris et al.*, 2016], while a four-stage model for the preparation of an EQ by means of its observable EM activity has been recently put forward [*Donner et al.*, 2015; *Potirakis et al.*, 2016, and references therein].

In this letter, we report the recording, with a sampling rate of 1 sample/s, of MHz EME signals with critical characteristics of a second order phase transition in equilibrium before the EQs which happened in the region of the eastern Aegean Sea, between the Greek Islands of Lesvos (6-7, 12 February 2017 and 12 June 2017), and Kos (20 July 2017) and the Turkish Asia Minor coastline. Moreover, we report tricritical characteristics revealed in the kHz EME recorded prior to the 12 June main event, while tricritical characteristics present both MHz and kHz EME prior to the 20 July event. Note that, as expected, all trictitical behaviors appear later than the corresponding MHz EME critical behavior. The transition from the critical to tricritical phase means that the system has departed from the equilibrium state. Furthermore, it has to be mentioned that no kHz signals of clearly high organization and persistency was possible to be validated, probably due to the fact that all the aforementioned EQs happened in the Sea or very close to it. We will further comment on this fact in the discussion-conclusion Section.

The analysis of the specific EME time-series was performed using the method of critical fluctuations (MCF) [*Contoyiannis and Diakonos*, 2000; *Contoyiannis et al.*, 2002, 2013, 2015; *Potirakis et al.*, 2016]. The analysis reveals that first in the timeline appear critical features in the MHz EME, implying that the possibly related underlying fracture process involved in the preparation of the main shock is at critical state. The presence of the "critical point" during which any two active parts of the system are highly correlated even at arbitrarily long distances, in other words when "everything depends on everything else", is consistent with the view that the EQ preparation process during the period that the MHz EME precursory signals are emitted is a spatially extensive process. It is noted that, according to the aforementioned four-stage model [*Potirakis et al.*, 2016, and references therein], the pre-seismic critical MHz EM emission is considered to originate during the fracture of the part of the Earth's crust that is characterized by highly heterogeneity. During this phase, the fracture is non-directional and spans over a large area that surrounds the family of large high-strength entities (asperities) distributed along the main fault sustaining the system. Note that for an EQ of magnitude ~6 the corresponding fracture process extends to a radius of ~120km [*Bowman et al.*, 1998]. Thus, during this phase the fracture process is extended up to the land of the neighboring islands. For two of the EQs of interest, the 12 June and 20 July events, the analysis also reveals that next in the timeline, after the critical MHz EME, appear tricritical EME (kHz in the first case, while both MHz and kHz in the second one). Tricritical dynamics signify the departure from critical state, implying that the possibly related underlying fracture process involved in the preparation of the main shock evolves from the highly symmetrical and spatially expanded phase to a low symmetry, spatially focused on a preferred direction (along the fault) phase [*Contoyiannis et al.*, 2015; *Potirakis et al.*, 2016].

## 2  Data analysis method

The analysis of the recorded data was performed using the method of critical fluctuations (MCF) [*Contoyiannis and Diakonos*, 2000; *Contoyiannis et al.*, 2002]. Detailed descriptions of all the involved calculations can be found elsewhere [*Contoyiannis et al.*, 2013, 2015; *Potirakis et al.*, 2016] and therefore are omitted here for the sake of brevity and focus on the findings. However, a general description of the employed method follows.

MCF was proposed for the analysis of critical fluctuations in the observables of systems that undergo a continuous (second-order) phase transition [*Contoyiannis and Diakonos*, 2000; *Contoyiannis et al.*, 2002]. It is based on the finding that the fluctuations of the order parameter, that characterizes successive configurations of critical systems at equilibrium, obey a dynamical law of intermittency of an 1D nonlinear map form. The MCF is applied to stationary time windows of statistically adequate length, for which the distribution of the of waiting times $l$ (laminar lengths) of fluctuations in a properly defined laminar region is fitted by a function $f(l) \propto l^{-p_2} e^{-p_3 l}$. The criteria for criticality are $p_2 > 1$ and $p_3 \approx 0$ [*Contoyiannis and Diakonos*, 2000; *Contoyiannis et al.*, 2002]. In that case the system is characterized by intermittent dynamics, since the distribution follows power-law decay [*Schuster*, 1998]. On the other hand, in the case of a system governed by noncritical dynamics the corresponding distribution follows an exponential decay, rather than a power-law one [*Contoyiannis et al.*, 2004b]. A system in critical dynamics may depart from this high symmetry phase towards a low symmetry (highly localized along a preferred direction) either by the so-called "symmetry-breaking" phenomenon [*Contoyiannis et al.*, 2004a], or by means of a tricritical crossover [*Contoyiannis et al.*, 2015]. The so-called "tricritical point" is the point in the phase diagram of the system at which the two basic kinds of phase transition (second-order phase transition and first-order phase transition) meet. Both these ways of departure from critical state can be detected by using MCF. Symmetry breaking is characterized by marginal presence of power-law distribution, which indicates that the system's state is still close to the critical point. Using MCF, this is identified by the appearance of an extremely narrow range of laminar regions (often just one region) for which criticality conditions are still satisfied, after the appearance of a clear critical window [*Contoyiannis et al.*, 2004a]. Tricritical dynamics are identified in terms of MCF by the appearance of the combination of exponents $p_2 < 1$ and $p_3 \approx 0$ [*Contoyiannis et al.*, 2015]. The MCF has been applied to a variety of dynamical systems, including thermal (e.g., 3D Ising) [*Contoyiannis et al.*, 2002], geophysical [*Contoyiannis et al.*, 2004a, 2005, 2010, 2013, 2015; *Contoyiannis and Eftaxias* 2008; *Potirakis et al.*, 2016], biological (electro-cardiac signals) [*Contoyiannis et al.*, 2004b; *Contoyiannis et al.*, 2013], economic [*Ozun et al.*, 2014] and electronic systems [*Potirakis et al.*, 2017].

## 3  Analysis results

As already mentioned, in this letter we present results related to the recent EQs which happened in the region of the eastern Aegean Sea, between the Greek Islands of Lesvos (6-7, 12 February 2017 and 12 June 2017), and Kos (20 July 2017) and the Turkish Asia Minor coastline. Specifically, we refer to the shallow EQs shown in Table 1, which we divide into four cases hereafter referred to as: (a) "Lesvos Feb. 6-7" (3 EQs, ~$M_L$5, occurred on 6-7 February 2017), (b) "Lesvos Feb. 12" (1 EQ, ~$M_L$5, occurred on 12 February 2017), (c) "Lesvos June 12" (1 EQ, ~$M_L$6, occurred on 12 June 2017), and (c) "Kos July 20" (1 EQ, ~$M_L$6, occurred on 20 July 2017).

Contoyiannis et al.    Preseismic electromagnetic emissions related to recent eastern Aegean Sea EQs.    p. 4

**Table 1**. Detailed information about the EQs of interest (date/time refer to UT).

| Date | Time | Longitude | Latitude | Depth (km) | $M_L$ |
|---|---|---|---|---|---|
| **Lesvos Feb. 6-7** | | | | | |
| 2017 FEB 6 | 03:51:41.6 | 39.51 | 26.10 | 14 | 5.0 |
| 2017 FEB 6 | 10:58:02.5 | 39.51 | 26.12 | 15 | 5.1 |
| 2017 FEB 7 | 02 24 04.6 | 39.49 | 26.11 | 16 | 5.2 |
| **Lesvos Feb. 12** | | | | | |
| 2017 FEB 12 | 13:48:16.5 | 39.51 | 26.14 | 12 | 4.9 |
| **Lesvos June 12** | | | | | |
| 2017 JUN 12 | 12:28:38.2 | 38.84 | 26.36 | 12 | 6.1 |
| **Kos July 20** | | | | | |
| 2017 JUL 20 | 22:31:11.7 | 36.96 | 27.43 | 10 | 6.2 |

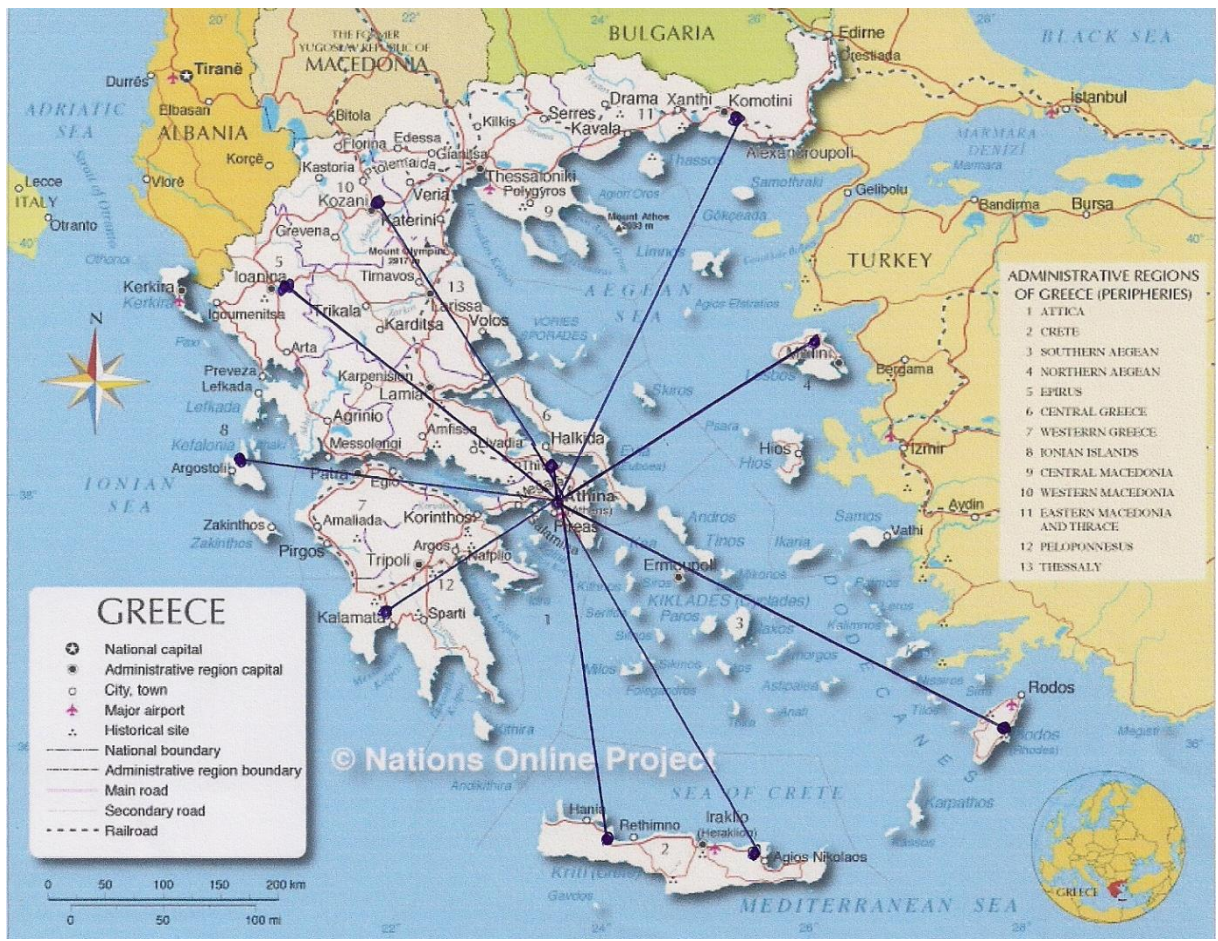

**Fig. 1.** The 10 currently operating remote sensing stations of the telemetric network for the recording of MHz and kHz EME in Greece.

**Table 2.** Location of the currently operating remote EME stations

| No. | Code Name | Station | Latitude | Longitude | Altitude (above Sea level) |
|---|---|---|---|---|---|
| 1 | J | Ioannina | 39.6561 ° N | 20.8487 ° E | 526 m |
| 2 | H | Atalandi | 38.6495 ° N | 22.9988 ° E | 185 m |
| 3 | F | Balsamata, Cephalonia Island | 38.1768 ° N | 20.5886 ° E | 402 m |
| 4 | O | Ithomi, Kalamata | 37.1787 ° N | 21.9252 ° E | 423 m |
| 5 | K | Kozani | 40.3033 ° N | 21.7820 ° E | 791 m |
| 6 | E | Neapoli, Crete Island | 35.2613 ° N | 25.6103 ° E | 288 m |
| 7 | V | Vamos, Crete Island | 35.4070 ° N | 24.1997 ° E | 225 m |
| 8 | A | Archangelos, Rhodes Island | 36.2135 ° N | 29.1212 ° E | 148 m |
| 9 | T | Komotini | 41.1450 ° N | 25.5355 ° E | 116 m |
| 10 | M | Agia Paraskevi, Lesvos Island | 39.2456 ° N | 26.2649 ° E | 130 m |

The hereafter presented signals were recorded in specific remote stations of our telemetric network spanning across Greece (Fig. 1, Table 2). In the following we present the results for each one of the aforementioned four cases, in chronological order.

### 3.1 Lesvos Feb. 6-7 case

The MCF analysis results reveal that MHz EME recorded at the measurement station located on the island of Lesvos (Fig. 1, Table 2) present criticality features for a continuous period of at least one day on 02 February 2017, 4-5 days before the 6-7 February EQs (see Table 1). Fig. 2a shows an one-day-long "critical window" (CW) for which criticality conditions are satisfied. CWs are time intervals of the MHz EME signals presenting features analogous to the critical point of a second order phase transition [*Contoyiannis et al.*, 2005]. The extraordinary continuous duration for which the system stays in criticality for EQs with magnitude approximately 5 (the usual duration is a few hours) may correspond to the simultaneous preparation process of the three EQs in practically the same heterogeneous region surrounding the finally activated neighboring faults.

The main steps of the MCF analysis [e.g., *Contoyiannis et al.*, 2013] on the specific time-series are shown in Figs. 2b-d. First, a distribution of the amplitude values of the analyzed signal was obtained (Fig. 2b) from which, using the method of turning points [*Pingel et al.*, 1999], a fixed-point, that is the start of laminar regions, $\phi_o$ was determined. Fig. 2c portrays an example of the obtained laminar distributions for a specific end point $\phi_l$, that is the distribution of waiting times, referred to as laminar lengths $l$, between the fixed-point $\phi_o$ and the end point $\phi_l$, as well as the fitted function $f(l) \propto l^{-p_2} e^{-p_3 l}$. Finally, Fig. 2d shows the obtained plot of the $p_2$, $p_3$ exponents vs. $\phi_l$. From Fig. 2d it is apparent that the criticality conditions, $p_2 > 1$ and $p_3 \approx 0$, are satisfied for a wide range of end points $\phi_l$, revealing the power-law decay feature of the time-series which indicates that the system is characterized by intermittent dynamics; in other words, the MHz time-series excerpt of Fig. 2a is indeed a CW.

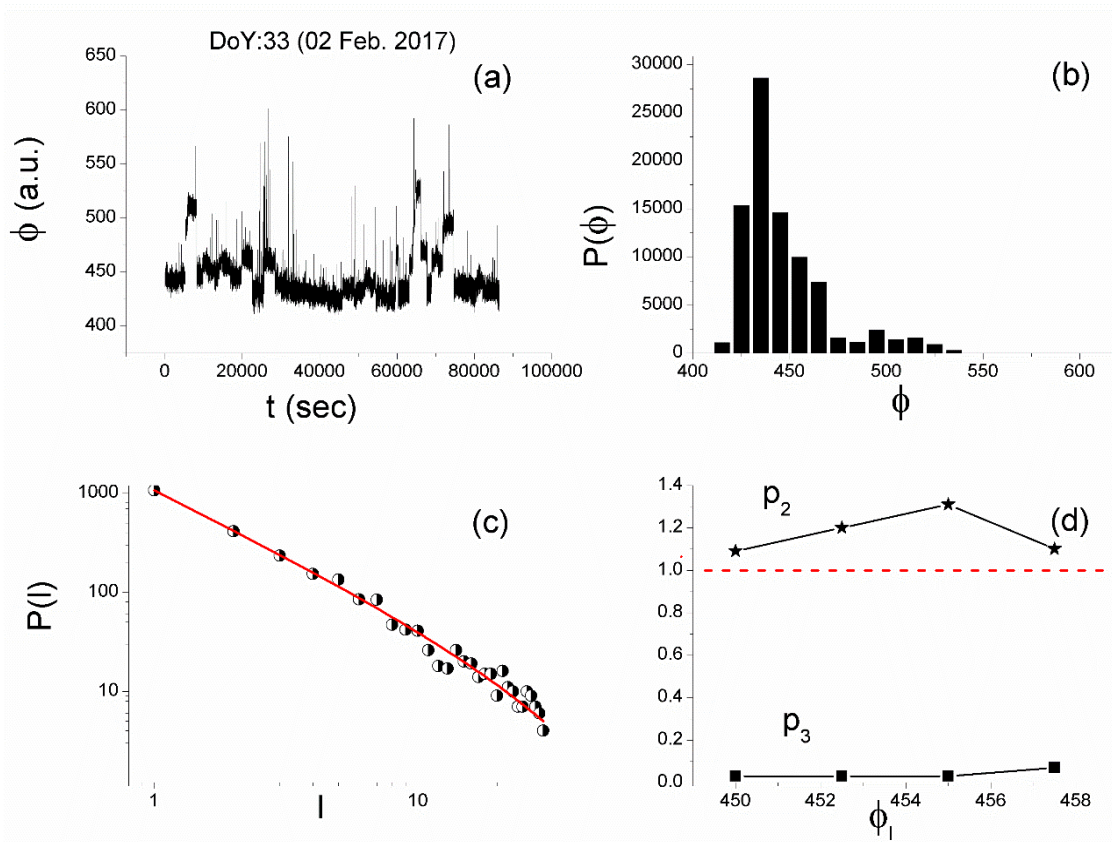

**Fig. 2**. (a) The 86400 samples long critical window of the MHz EME that was recorded before the Levos Feb. 6-7 EQs at the Lesvos island station (see also Fig. 1, Table 2). (b) Amplitude distribution of the signal of (a). (c) A representative example of laminar distribution and the involved fitting. The solid line corresponds to the fitted function (cf. to text in Sec. 2). (d) The obtained exponents $p_2$, $p_3$ vs. different values of the end of laminar region $\phi_l$. The horizontal dashed line indicates the critical limit ($p_2 = 1$). Critical behavior is obvious.

### 3.2 Lesvos Feb. 12 case

The MCF analysis results reveal that MHz EME recorded at the measurement station located on the island of Lesvos (Fig. 1, Table 2) present criticality features for a 5 h long excerpt on 07 February 2017 (Fig. 3), 5 days before the 12 February EQ. Note that the specific CW was detected ~ 10 h after the third EQ of the Lesvos Feb. 6-7 discussed in Sec. 3.1 (see Table 1).

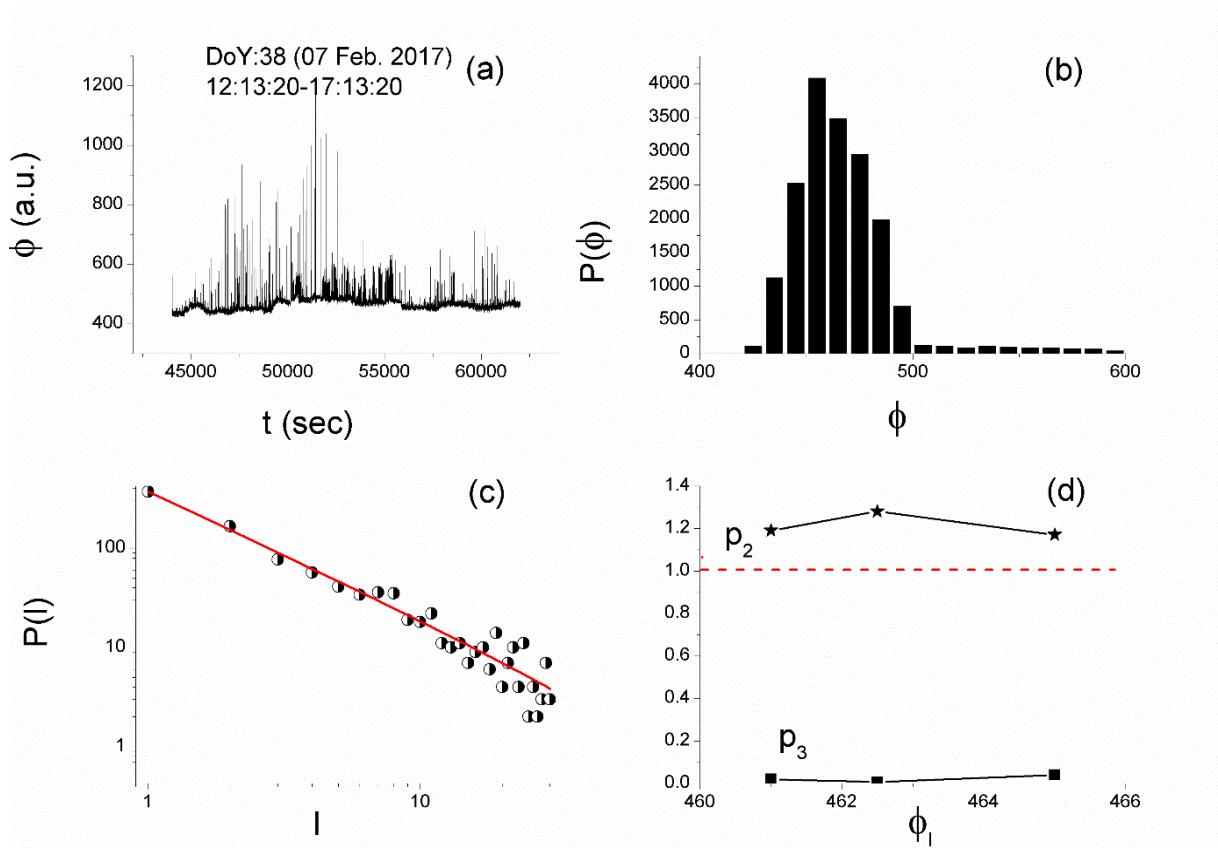

**Fig. 3**. (a) A 5 h (18000 samples) long excerpt of the MHz EME time series recorded before the Levos Feb. 12 EQ at the Lesvos island station (see also Fig. 1, Table 2). (b)-(d) similar to Fig. 2. Critical behavior is obvious.

### 3.3 Lesvos June 12 case

The MCF analysis results reveal that MHz EME recorded at the measurement station located on the island of Lesvos (Fig. 1, Table 2) present criticality features for a ~1.7 h long (6000 samples) excerpt on 01 June 2017 (Fig. 4), ~11.5 days before the 12 June EQ (see Table 1).

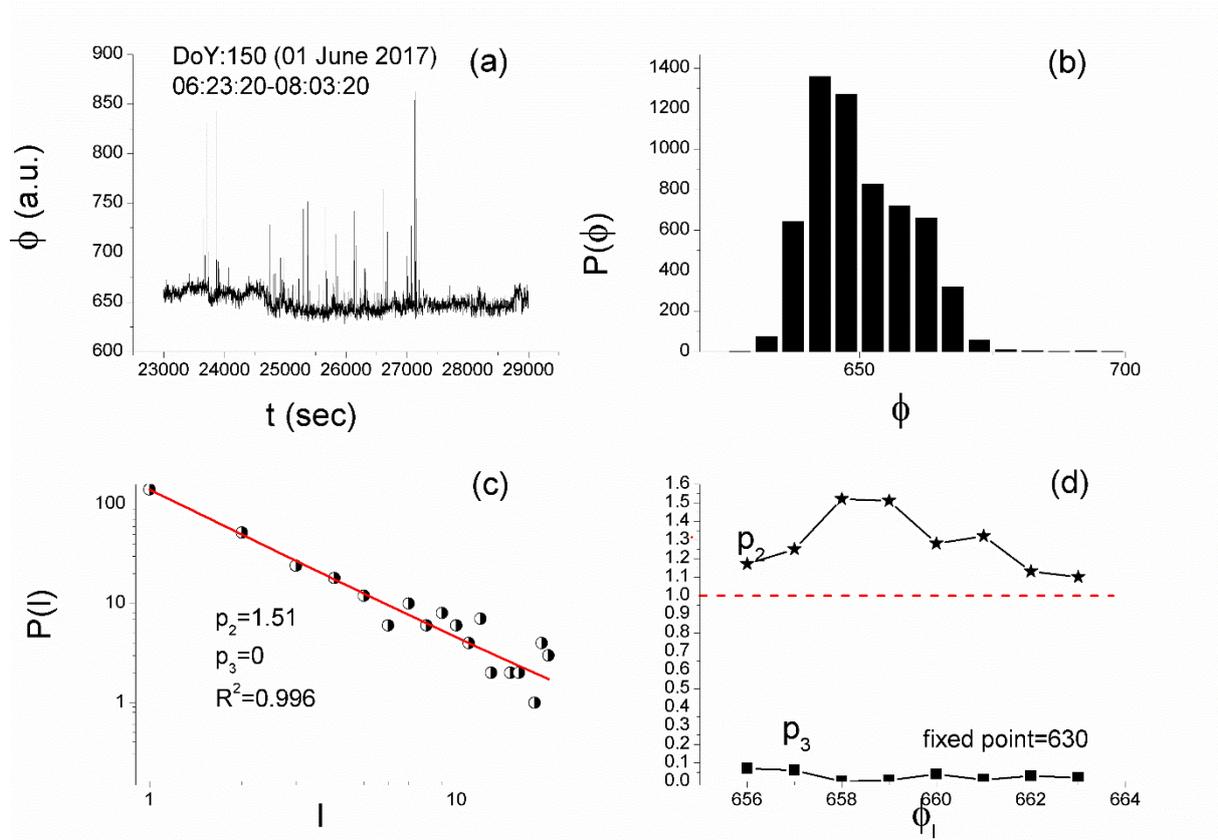

**Fig. 4**. (a) A ~1.7 h (6000 samples) long excerpt of the MHz EME time series recorded before the Levos June 12 EQ at the Lesvos island station (see also Fig. 1, Table 2). (b)-(d) similar to Fig. 2. Critical behavior is obvious.

Importantly, a few days after the appearance of the CW of Fig. 4a, tricritical behavior was found in the kHz EME recordings of the same station (both at the 3kHz and 10kHz) 7-6 days before the Lesvos June 12 EQ. As it has been mentioned, this result indicates that the under-fracture system has departed from the condition of equilibrium. An example for 3kHz is shown in Fig. 5. From Fig. 5d it is obvious that the tricriticality conditions $p_2 < 1$ and $p_3 \approx 0$ are satisfied for a wide range of end points $\phi_l$,

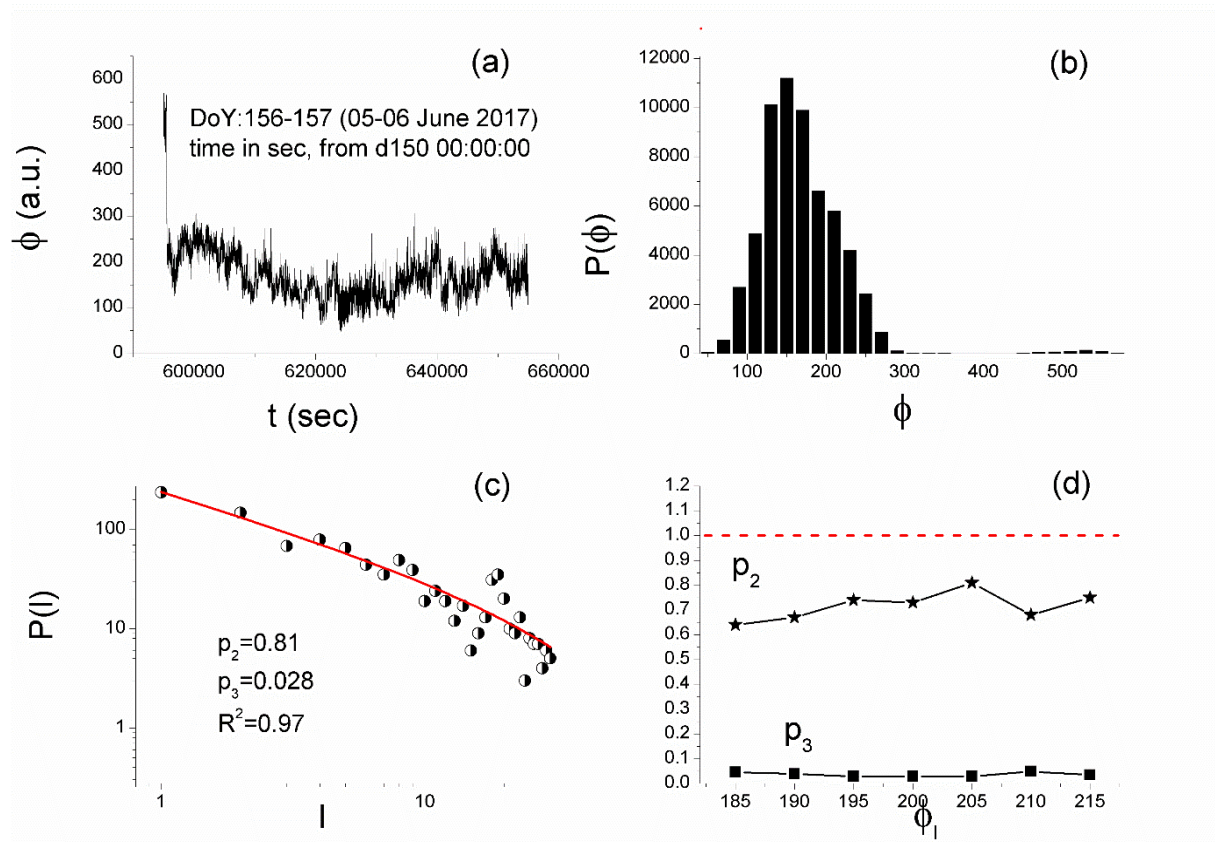

**Fig. 5**. (a) A ~16.7 h (60000 samples) long excerpt of the kHz EME (3kHz) time series recorded before the Levos June 12 EQ at the Lesvos island station (see also Fig. 1, Table 2). (b)-(d) similar to Fig. 2. Tricritical behavior is obvious.

### 3.4 Kos July 20 case

The MCF analysis results reveal that MHz EME recorded at the measurement station located on the island of Rhodes (Fig. 1, Table 2) present criticality features for a continuous period of at least one day on 19 July 2017 (Fig. 6), just 1 day before the 20 July Kos EQ (see Table 1).

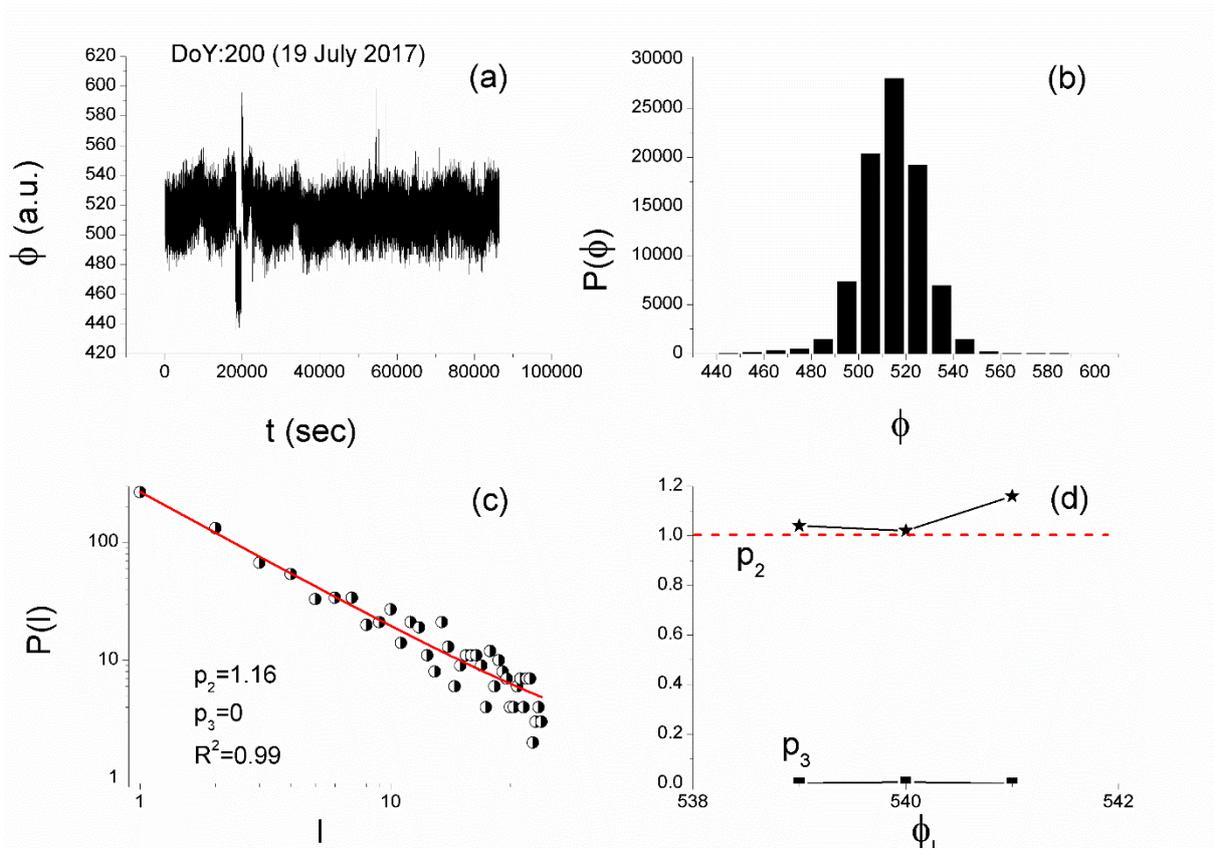

**Fig. 6**. (a) An one day long excerpt of the MHz EME time series recorded before the Kos July 20 EQ at the Rhodes island station (see also Fig. 1, Table 2). (b)-(d) similar to Fig. 2. Critical behavior is obvious.

A few hours later, and only ~3 h before the main shock, the same station recorded a 2.5 h long tricritical MHz EME, as shown in Fig. 7. Tricritical behavior is obvious, although for just one laminar region, it seems that critical dynamics leftovers are present.

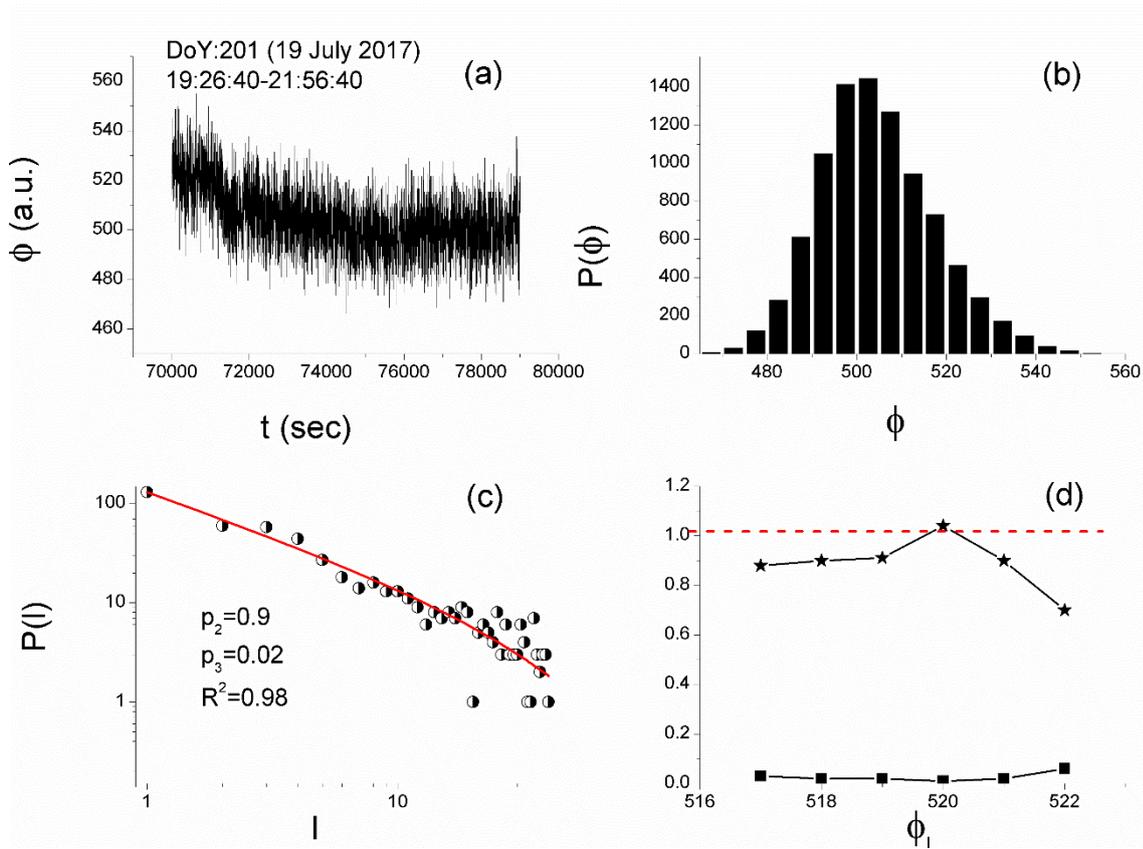

**Fig. 7**. (a) A 2.5 h (9000 samples) long excerpt of the MHz EME time series recorded before the Kos July 20 EQ at the Rhodes island station (see also Fig. 1, Table 2). (b)-(d) similar to Fig. 2. Tricritical behavior is obvious.

The whole process seems to be "compressed" in time, very close to the main shock occurrence. Fig. 8 shows that no indications for criticality appeared before 19 July 2017.

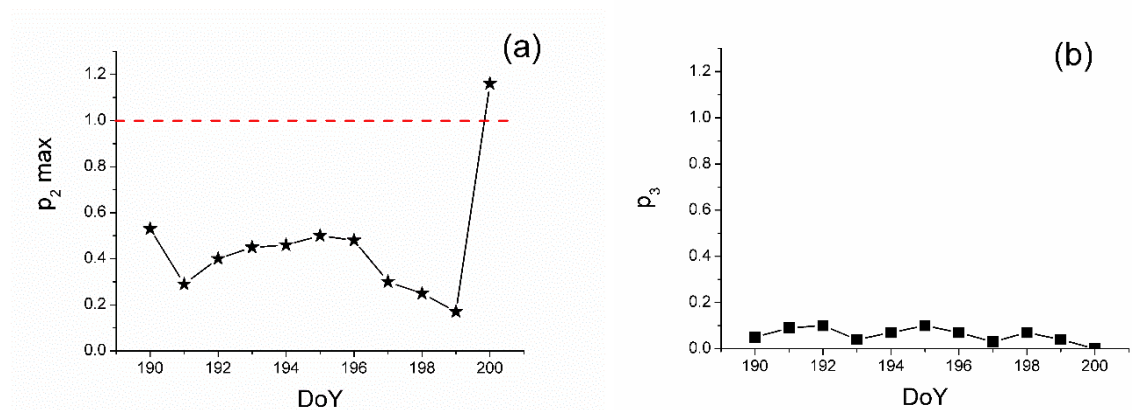

**Fig. 8**. Change of the exponents MCF analysis exponents over time for the MHz time series recorded at Rhodos station (a) $p_2$, (b) $p_3$. Criticality appears no earlier than 19 July.

## 4  Discussion - Conclusions

Based on the method of critical fluctuations, we have shown that the recent seismic events that occurred in Greece also shown successively critical-tricritical precursory behavior by means of the fractured-induced EME. These two successively appearing behaviors correspond to the first and second stage of our proposed four-stage model of EQ preparation process in terms of fracture induced EME [*Potirakis et al.*, 2016, and references therein]. The observed EM silence before the EQ occurrence corresponds to the fourth stage of this model. We note that crucial very strong avalanche like precursory kHz EME signals, which are recorded in the tail of the preseismic kHz EM emission in the case of shallow EQs that happen inland (or near coast-line) and correspond to the third stage of the model, were not recorded prior to the seismic events under study. It has been suggested that the lounge of the kHz EM activity shows the fracture of asperities sustaining the main fault corresponding to the third stage of the four-stage model. In the cases under study the fracture process during this stage has been confined to a narrow area, i.e., along the main fault in a submerged area. This situation by itself justifies the observed absence. On the other hand, in terms of percolation theory, the "hydraulic threshold", $x_c$, during which the transition impermeable–permeable occurs, as well as the "mechanical" or "damage threshold", $x_m$ $(x_m > x_c)$, during which the infinite cluster (IC) is formed and the solid disintegrates, precede the shear displacement along the fault plane (the EQ), which means that many transport properties are activated before the main event [*Eftaxias et al.*, 2013, and references therein]. This fact, combined by the governing role of local hydraulic conditions on the water injection-induced fracture behavior in rocks concerning the mechanical properties, fracture nucleation and the geometry of the shear fracture zone [*Li et al.*, 2016], further enhances the absence of the final strong pulse like kHz EME. We note that due to the crucial character of this emission, an austere set of criteria have been established to characterize such a recorded kHz anomaly as a seismogenic one [*Eftaxias and Potirakis*, 2013; *Eftaxias et al.*, 2013]. The multidisciplinary analysis of the kHz records before the under study seismic events in terms of these criteria verified the absence of the last emerged part of the fracture induced kHz EME corresponding to the fracture of asperities.

As it has been repeatedly pointed out in our works, our view is that such observations and the associated analyses offer valuable information for the comprehension of the Earth system processes that take place prior to the occurrence of a significant EQ. As it is known a large number of other precursory phenomena are also observed, both by ground and satellite stations, prior to significant EQs. Only a combined evaluation of our observations with other well documented precursory phenomena could possibly render our observations useful for a reliable short-term forecast solution. In the cases under study this requirement was not fulfilled.